# Human-Robo-Advisor Collaboration in Decision-Making: Evidence from a Multiphase Mixed Methods Experimental Study

Hasan Mahmud[a*], A.K.M. Najmul Islam[c], Satish Krishnan[b,c]

[a] *Rochester Institute of Technology, Rochester, New York, USA, hxmdfp@rit.edu*
[b] Information Systems Area, Indian Institute of Management Kozhikode, India, *satishk@iimk.ac.in*
[c] *Software Engineering Department, LUT University, Lappeenranta, Finland, najmul.islam@lut.fi*

**Abstract**

Robo-advisors (RAs) are cost-effective, bias-resistant alternatives to human financial advisors, yet adoption remains limited. While prior research has examined user interactions with RAs, less is known about how individuals interpret RA roles and integrate their advice into decision-making. To address this gap, this study employs a multiphase mixed methods design integrating a behavioral experiment (N = 334), thematic analysis, and follow-up quantitative testing. Findings suggest that people tend to rely on RAs, with reliance shaped by information about RA performance and the framing of advice as gains or losses. Thematic analysis reveals three RA roles in decision-making and four user types, each reflecting distinct patterns of advice integration. In addition, a 2×2 typology categorizes antecedents of acceptance into enablers and inhibitors at both the individual and algorithmic levels. By combining behavioral, interpretive, and confirmatory evidence, this study advances understanding of human–RA collaboration and provides actionable insights for designing more trustworthy and adaptive RA systems.

## 1. Introduction

Financial decision-making is inherently complex, prompting individual investors to seek professional guidance from human financial advisors [1]. However, such advisors are often criticized for conflict of interest [2], high fees [3], and emotional biases [4], leading many investors to seek alternatives. In response, algorithmic advancements have introduced automated platforms, known as robo-advisors (RAs), which provide personalized investment guidance and portfolio management [1,5,6]. RAs are accessible, cost-effective, and can reduce human error and behavioral bias [5,7,8]. Despite these advantages, adoption remains limited, with many individuals still turning to informal advice sources such as online forums or social media. As of 2022, less than 1% of global assets are managed by RAs [9,10], far below those managed by leading firms such as BlackRock or Vanguard [11]. The reasons behind this slow uptake remain underexplored [4,12].

Financial decisions involve high stakes, requiring long-term planning and complex trade-offs under uncertainty. Unlike lower-stakes domains such as entertainment or online shopping, financial contexts demand more cautious and deliberative engagement with decision-support tools like RAs. This cautious stance heightens users' sensitivity to the credibility and functionality of RAs, leading them to scrutinize the role and trustworthiness of their advice. Given this closer scrutiny, it is important to examine why individuals accept or reject RA advice in order to better understand human–RA collaboration.

Human–RA collaboration can be defined as an interactive process in which a human and an RA jointly participate in decision-making [13]. Unlike fully automated or purely human approaches, this collaboration leverages the complementary strengths of both agents. Its success depends on how effectively these two entities interact—a process shaped by technical factors (e.g., transparency, advice framing, reliability) and human factors related to cognition (e.g., self-confidence, reasoning, interpretation), affect (e.g., trust), and behavior (e.g., advice integration). Research across disciplines such as Information Systems (IS), Finance, Accounting, and Marketing has made important strides in examining these factors [2,5,7,8,14,15]. However, several critical gaps exist. *First*, the influence of two key RA design elements—performance information (e.g., historical accuracy rates) and prediction framing (optimistic vs. pessimistic)—on user reliance has received limited empirical attention. Performance information, as a form of algorithmic transparency, can signal competence and enhance trust [16]. However, prior findings are mixed: some studies report increased reliance when performance information is disclosed [17–19], while others observe persistent algorithm aversion despite superior outcomes [20,21]. Similarly, prediction framing, whether the RA presents an optimistic or pessimistic outlook, can influence risk perception and user responses. Consistent with Prospect Theory, individuals tend to weigh losses more heavily than gains [22], making pessimistic framing potentially more persuasive. Yet empirical evidence is fragmented: some studies find optimistic outlooks encourage reliance [23], while others report no significant effects [18]. These inconsistencies highlight the need for further empirical investigation.

*Second,* little is known about the roles RAs play during decision-making. While prior research has focused on whether users accept or reject algorithmic input, it often overlooks the interpretive lens through which users make sense of it. Some may view the RA as confirming their initial judgment, others as prompting reflection or reassessment, and still others as offering complementary insights. These role interpretations shape not only reliance behavior but also how users perceive the RA's function and value in the decision process. Understanding such subjective interpretations is essential for designing RAs that align with user expectations and foster effective human–RA collaboration.

*Third,* existing studies lack empirical evidence on how individuals integrate algorithmic advice into their decisions. Most prior work focuses on intended acceptance or self-reported trust [4–6,8,12,14,24–26], which may not accurately reflect actual behavioral reliance. Yet understanding behavioral reliance is essential, as users may express trust without incorporating algorithmic input or rely on it despite skepticism. Even among those who rely on RA advice, integration patterns vary significantly: some users closely follow the advice, others adjust partially, and some disregard it altogether. These variations reflect differing levels of trust and engagement. Therefore, it is critical to examine how users adjust their decisions in response to RA input. Such behavioral evidence can provide a more accurate and actionable understanding of human–RA collaboration than intention-based measures alone.

*Finally*, most RA studies rely on theory-driven quantitative approaches that test predefined constructs such as perceived usefulness, ease of use, or risk and their relationships [12,24,25,27,28]. While these models offer structured explanations, they often overlook important aspects of user experience that emerge organically during interaction. An exploratory qualitative approach is necessary to uncover these underexplored but meaningful dimensions of human–RA collaboration.

To address these gaps, this study poses three central questions: (1) To what extent do individuals rely on RA advice, and how do performance information and prediction framing influence this reliance? (2) How do individuals interpret RA roles and integrate its advice into their decision-making? (3) What user- and RA-specific factors shape human-RA collaboration? Answering these questions requires an approach that captures both behavioral reliance and subjective experience. Accordingly, this study employs a multiphase mixed methods design. Phase One quantitatively measures behavioral reliance using the Weight of Advice (WOA) metric. Introduced in the Judge–Advisor System (JAS) paradigm, WOA captures how individuals revise their judgments after receiving external advice [29]. This setup enables a controlled examination of users' responses to RA advice under varying conditions of performance information and prediction framing (RQ1). Phase Two employs qualitative inquiry to explore how users interpret RA roles and integrate its advice (RQ2) and to identify individual- and RA-level factors shaping collaboration (RQ3). Phase Three conducts follow-up quantitative testing to assess whether users' interpretation styles and integration behaviors correspond to significant differences in behavioral reliance. This mixed-methods approach allows us to examine key factors shaping human–RA collaboration, including performance information, prediction framing, interpretation of RA roles, trust, and patterns of advice integration. By integrating experimental evidence with qualitative insights, this design informs a holistic framework for understanding human–RA collaboration.

## 2. Theoretical Background

### 2.1 RAs and Their Acceptance

RAs are automated financial platforms that use algorithms to deliver personalized investment recommendations and manage portfolios [1,5,6]. They typically begin by assessing an investor's risk tolerance and financial goals, often using online questionnaires, and then either provide tailored advice or manage investments autonomously [1,24,30].

Despite their advantages, including convenience, personalization, and low cost, RA adoption remains limited [8]. To better understand this gap, researchers have drawn on theoretical frameworks such as the Technology Acceptance Model, Value-Based Adoption Model, Diffusion of Innovation, Information Search Model, Dual Process Theory, Self-Service Technology Adoption Theory, Theory of Planned Behavior, and Social Response Theory [6,8,12,15,24,25,27]. Empirical studies based on these models identify various human and algorithm-specific factors influencing RA acceptance. On the human side, demographic traits such as age and financial position play a role. For example, younger individuals and those with higher investable assets are more likely to adopt RAs, while gender has shown limited influence [12]. Psychological traits such as extroversion and risk tolerance are positively associated with adoption, as is financial knowledge [4]. In contrast, habitual resistance to algorithmic decision-making can act as a barrier [3]. Additionally, perceptions of the technology, such as perceived usefulness, ease of use, and benefits, are consistently linked to higher adoption intentions, whereas perceived risks tend to reduce them [15,24]. On the algorithm side, design features significantly influence user trust and acceptance. For example, anthropomorphic interfaces can increase perceived social presence of the RA, fostering greater user trust [25]. Similarly, non-human-like names have been found to generate more favorable responses [14]. Other vital features include personalization, transparency, reliability, and data privacy, each of which has been shown to enhance user acceptance [5,30]. In addition, an active social media presence of RAs can increase familiarity and reduce perceived risk [24]. More recent findings suggest that interpretability and interactivity strengthen users' intention to rely on RAs [6].

Together, these studies have advanced our understanding of RA adoption by identifying key predictors of user intention to accept RA. However, much of this literature centers on intended acceptance or self-reported trust, offering limited insight into how individuals actually rely on RA advice and interpret the roles RAs play. Moreover, prior studies often examine enablers or inhibitors of RA reliance in isolation [31], rather than considering both dimensions concurrently. Technical design elements such as performance information and prediction framing have also received limited empirical attention, despite their potential to shape user reliance. These limitations point to the need for research that captures behavioral outcomes and explores the cognitive and technical dimensions of human–RA collaboration.

## 2.2 JAS and Behavioral Trust

A key objective of this study is to capture actual behavioral reliance on RA advice. To this end, we adopt the JAS as the guiding framework [29]. The JAS has been widely applied in the advice-taking literature [20,32] to examine how individuals revise their judgments after receiving external input. For example, Ning et al. [33] used JAS to investigate how transparency features influence algorithmic advice use, while Logg et al. [32] applied it to compare reliance on algorithmic versus human recommendations. Within this paradigm, participants (judges) make an initial estimate (e.g., predicting a financial index), view the advisor's estimate (e.g., RA's estimate), and then submit a final estimate. They are free to fully adopt, partially adjust, or disregard the advisor's input. Behavioral reliance on advice is quantified using the WOA metric:

$$\text{WOA} = \frac{\text{Adjusted Estimate} - \text{Initial Estimate}}{\text{Advisor Estimate} - \text{Initial Estimate}}$$

Where:
- Initial Estimate = Participant's prediction before viewing the advisor's estimate
- Advisor Estimate = Prediction provided by the advisor
- Adjusted Estimate = Participant's final prediction after seeing the advisor's estimate

WOA theoretically ranges from 0 to 1. A WOA of 1 indicates complete acceptance (full alignment with the RA), while a WOA of 0 reflects full rejection (no change from the initial estimate). Values between 0 and 1 indicate partial acceptance. However, WOA values can fall outside this range when participants adjust beyond the RA's estimate. For example, if the initial estimate is 4500, the RA suggests 4216, and the participant revises to 4000, the resulting WOA exceeds 1, indicating over-adjustment. To address such cases, prior studies have either retained values outside the theoretical range [20] or Winsorized them, setting values below 0 to 0 and above 1 to 1 [32,34]. This approach minimizes the influence of outliers while preserving the full dataset and improving the robustness of statistical analyses [35]. Following this convention, the present study adopts Winsorization to ensure consistency, comparability, and interpretability in measuring behavioral reliance [32,33].

## 3. Hypotheses Development

### 3.1 Reliance on RA

RAs offer several advantages that make them attractive to investors, with cost-efficiency being a primary benefit. While human financial advisors often charge fees ranging from 2% to 20%, RAs typically operate at a much lower cost, around 0.25%, and require minimal opening balances, often one-hundredth of those required by traditional advisors [5,6,24,25,36]. Beyond affordability, RAs provide continuous access via digital platforms, allowing users to obtain investment advice anytime without in-person appointments [4,5,24,36]. Functionally, they reduce investment barriers by offering diversified portfolios [1], requiring limited financial expertise [6], and delivering real-time responses to market conditions. RAs also integrate both subjective inputs (e.g., user preferences) and objective data (e.g., market indices), enhancing the precision of recommendations [6,8]. Importantly, they are less susceptible to emotional bias and human error [4,8,27]. Collectively, cost efficiency, accessibility, responsiveness, and perceived objectivity make RAs effective aids in investment decision-making. We therefore hypothesize:

*H1: Individuals are more likely to rely on RA advice when making investment decisions.*

## 3.2 Performance Information and Prediction Framing

Performance information refers to cues about how accurately an algorithm has performed in the past. According to *Signaling Theory*, such cues help reduce informational asymmetry and guide judgment when direct evaluation is difficult [16]. This is particularly relevant for RAs, which often function as black boxes, offering limited transparency into their underlying logic or decision processes. As a result, users typically lack sufficient insight to assess the system's reliability directly. In such cases, performance information can serve as a proxy for competence, enhancing perceived credibility and fostering decision confidence. When users are informed that an RA has a strong track record, they are more likely to interpret it as a sign of trustworthiness, increasing their willingness to rely on the advice.

Furthermore, *Cognitive Load Theory* posits that individuals have limited cognitive capacity for processing complex information [37]. When assessing an RA's reliability, performance cues can simplify decision-making by acting as mental shortcuts, thereby reducing cognitive effort. Instead of analyzing the RA's internal logic or technical details, users rely on performance signals to judge its trustworthiness and decision quality. Taken together, Signaling Theory and Cognitive Load Theory suggest that performance information functions both as a credibility signal and a cognitive aid, encouraging users to trust and follow RA advice more readily. Empirical research also supports this view, demonstrating that familiarity with an algorithm's prior success increases both acceptance and reliance [17–19]. Therefore, we hypothesize:

*H2a: Individuals are more likely to rely on RA advice when performance information is provided.*

Prediction framing refers to the directional emphasis of a forecast, specifically whether an RA presents an optimistic (gain-oriented) or pessimistic (loss-oriented) scenario. This framing can influence how users perceive and respond to the advice. *Prospect Theory* can explain this effect, which suggests that individuals are more sensitive to potential losses than to equivalent gains, a phenomenon known as loss aversion [22]. As a result, pessimistic forecasts may heighten perceived risk and prompt users to seek additional guidance, such as RA advice, to mitigate potential losses.

This effect is further reinforced by *negativity bias*, a cognitive tendency in which individuals perceive negative information as more salient, credible, and informative than positive cues [38,39]. As negative forecasts often deviate from expectations and signal caution, users may interpret them as more realistic and trustworthy [38,39]. In high-stakes contexts such as investment decision-making, where uncertainty is substantial, loss-oriented outlooks may appear more prudent and compelling. Taken together, these psychological mechanisms suggest that pessimistic framing can increase the perceived value of RA advice and promote greater reliance. Therefore, we hypothesize:

*H2b: Individuals are more likely to rely on RA advice when the prediction outlook is pessimistic rather than optimistic.*

## 4. Methodology

This study adopts a multiphase mixed methods experimental design to investigate individual reliance on RA advice. The research was conducted in three phases. The first phase involved a quantitative experiment to measure behavioral reliance by examining how participants adjusted their initial estimates after receiving input from an RA. The second phase was qualitative, involving thematic analysis of open-ended responses to uncover emergent concepts shaping RA acceptance. The third phase returned to a quantitative approach, aiming to validate selected findings from the qualitative analysis and reinforce the study's interpretive claims.

We designed an experiment to collect data where participants received RA advice under different conditions. Specifically, the design manipulated two key RA features: the presence or absence of performance information and the framing of prediction outlook (optimistic vs. pessimistic). This setup aimed to simulate real-world decision-making scenarios in which investors face uncertainty about the credibility of algorithmic advice and the direction of market forecasts.

## 4.1 Experimental Protocol and Pilot Testing

To simulate a realistic investment scenario, an experimental vignette (Appendix A) focused on the S&P 500 Index was developed, following Castelo et al. [17]. The vignette was pretested with eight experts from academia and industry, representing fields such as investment, accounting, software engineering, and business administration. Based on their feedback, revisions were made to improve clarity and contextual relevance.

To generate a credible RA forecast for the vignette, six financial forecasting experts from the pretesting panel were asked to estimate the index value one month ahead under each experimental condition. Their individual estimates were averaged following the "wisdom of crowds" approach [40,41], resulting in a mean absolute projected change of 258 points. This approach ensured that the RA's prediction was realistic and empirically grounded, rather than arbitrarily set. For experimental control, the same 258-point change from the current index value was applied in both the "optimistic" and "pessimistic" conditions. In the pessimistic condition, the RA predicted a drop of 258 points below the current index value, while in the optimistic condition, it predicted a rise of 258 points above that same value. This design allowed us to isolate the effect of prediction framing (optimistic vs. pessimistic) while controlling for the magnitude of the change.

In addition to prediction framing, the second key manipulation involved the RA's performance information. This was operationalized by setting the RA's historical accuracy at 80%, a value chosen to balance overly conservative and highly optimistic benchmarks. For comparison, Filiz et al. [42] used a 70% success rate in a similar vignette-based design. Likewise, Shashank et al. [43] developed a stock prediction algorithm with 72% accuracy. At the upper end, Sable et al. [44] reported an average accuracy of 97.36% in stock prediction algorithm. Positioned between such empirical benchmarks and consistent with prior experimental studies [45], the 80% figure was used as a credible and realistic manipulation of algorithmic performance.

After finalizing the experimental protocol, a pilot test was conducted with 25 participants recruited via Prolific, a crowdsourcing platform. As the responses demonstrated sufficient quality and no adjustments were necessary, we proceeded with full-scale data collection using the same structure and materials.

## 4.2 Participants

Participants were recruited through Prolific, widely used in IS research [46]. Prolific was selected for its superior data quality compared to alternatives [17]. To ensure a broad representation of user experiences, the study imposed no restrictions on participation except a minimum age of 18. A total of 366 participants completed the experiments. After applying quality checks, 334 valid responses, including those from the pilot test, were retained for analysis. The experiment lasted on average 9.52 minutes, and participants were compensated €2.50 in line with Prolific's compensation guidelines. The final sample consisted of 51.80% male participants, with a mean age of 29.36 years. The majority (96.71%) had at least an undergraduate degree, and 66.17% were employed either full- or part-time.

## 4.3 Procedure

The study employed a 2×2 between-subjects design: performance information (present vs. absent) and prediction framing (optimistic vs. pessimistic). Participants were randomly assigned to one of four experimental conditions (Table 1). The procedure began with informed consent, followed by a brief explanation of what an RA is, ensuring all participants had a basic understanding of the system. Participants then received detailed instructions and were tasked with predicting the S&P 500 Index value one month ahead, assuming the role of potential investors. To support their initial estimate, they were shown a one-year historical chart of the index, including key market statistics: the current index level, value one year ago, annual high and low points, and the largest monthly fluctuations. This information was intended to replicate a realistic decision environment and provide sufficient context for making an informed prediction. After submitting their initial estimate, participants were informed that a financial investment firm had developed an RA capable of making forecasts. Depending on the experimental condition, the RA was described as optimistic or pessimistic and presented a prediction that was either higher or lower than the current index level. In the performance information condition, participants were also shown a statement that the RA had an approximate accuracy rate of 80%. To reinforce these manipulations, the terms "optimistic," "pessimistic," and the accuracy statement were shown in bold.

**Table 1**: Distribution of Participants by Experimental Condition (N = 334)

| Condition | Optimistic outlook | Pessimistic outlook |
|---|---|---|
| Without performance information | 77 | 77 |
| With performance information | 79 | 101 |

Participants were then asked to submit a final estimate, with the option to revise their initial prediction. This was followed by an open-ended question: "Please briefly indicate why you did or did not change your initial estimate." These responses enabled a deeper exploration of participants' reasoning and how it shaped their reliance on RA advice. Fig.1 summarizes the procedure.

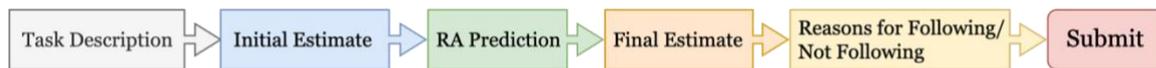

**Fig. 1.** Experiment Procedure

## 5. Findings

### 5.1 Findings of the First Quantitative Analysis

The quantitative inquiry revealed that approximately 67% of participants accepted the RA advice, while 33% did not (Fig. 2). A Chi-Square test indicated participants were significantly more likely to accept than reject RA advice ($\chi^2(1, N = 334) = 37.56$, $p<0.001$), supporting H1. Results from an independent samples t-test (Table 2) showed that exposure to performance information increased RA acceptance ($t = -2.006$, $p = 0.046$, $df = 332$), with a higher mean acceptance rate (0.481) compared to those without such information (0.398). Furthermore, the direction of the prediction outlook affects acceptance ($t = 2.197$, $p = 0.029$, $df = 332$), with pessimistic outlooks leading to higher acceptance (mean = 0.4848) than optimistic outlooks (mean = 0.3942). These results provide empirical support for H2a and H2b.

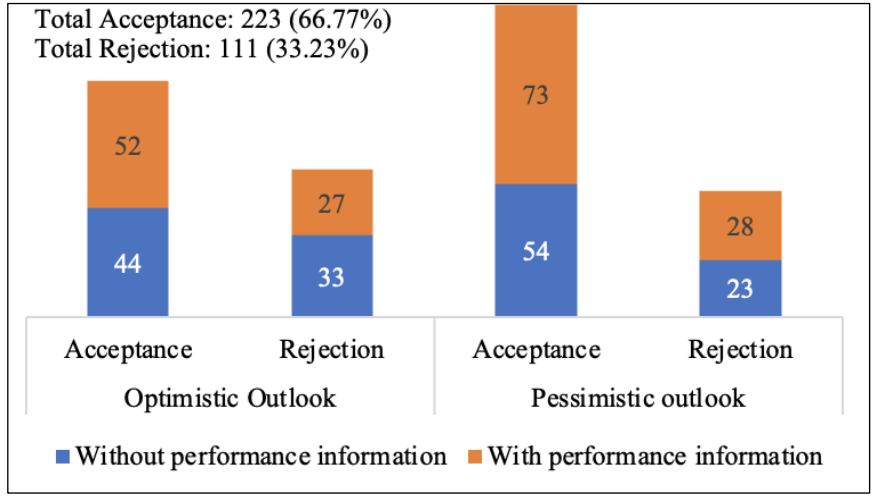

**Fig. 2.** Acceptance versus Rejection of RA

**Table 2**: Independent Samples T-Test Summary

| Comparison | Mean (Group 1) | Mean (Group 2) | Mean Difference | t-value | df | P-value | Cohen's d |
|---|---|---|---|---|---|---|---|
| With vs. Without Performance Information | 0.481 | 0.398 | -0.0829 | -2.006 | 332 | 0.046 | -0.22 |
| Pessimistic vs. Optimistic Outlook | 0.485 | 0.394 | 0.0906 | 2.197 | 332 | 0.029 | 0.241 |

*Note: Group 1 = "With Performance Information" / "Pessimistic Outlook"*
*Group 2 = "Without Performance Information" / "Optimistic Outlook"*

## 5.2 Findings of Qualitative Analysis

The quantitative analysis highlighted the measurable influence of RA advice on investor decision-making. To complement this, a qualitative inquiry was sought to explore why individuals respond differently to such advice. This inquiry was based on participants' responses to the open-ended question: *'Please briefly indicate why you did or did not change your initial estimate.'* This phase followed the Gioia methodology [47], a widely used framework in organizational and IS research [48] for structured thematic analysis. This method distinguishes three levels of abstraction: first-order concepts (participants' own words), second-order themes (researchers' interpretations), and aggregate dimensions (overarching theoretical categories). This structured approach ensures a transparent progression from raw data to theoretical insights, making it well-suited for examining underexplored or complex behavioral processes such as user engagement with algorithmic advice.

Following this approach, we reviewed all open-ended responses multiple times to familiarize ourselves with the data and understand participants' perspectives. We then independently coded the responses by assigning first-order codes to responses that conveyed meaningful rationales regarding RA use. These first-order codes captured participants' language as closely as possible. Next, through iterative refinement and discussion, we developed second-order themes that captured recurring patterns or shared meanings. Guided by our research question, we then grouped these themes into higher-order aggregate dimensions, reflecting how participants interpret RA roles and integrate RA advice and the underlying antecedents of RA acceptance (Fig. 3).

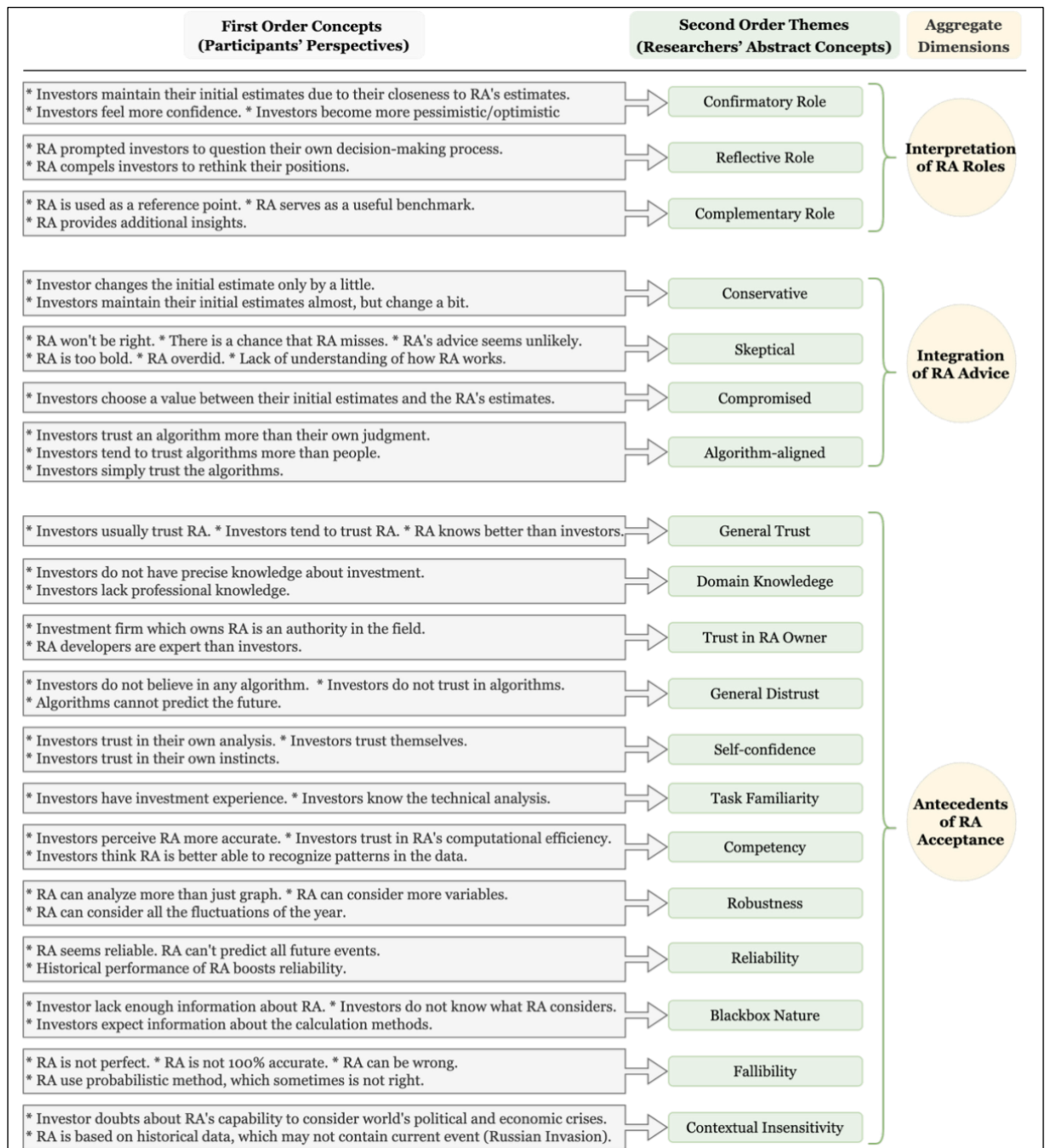

**Fig. 3.** Data structure showing interpretation of RA roles, integration of RA advice, and antecedents of RA acceptance.

### 5.2.1 Interpretation of RA Roles

Participants made sense of the RA's forecast in distinct ways, interpreting it as playing three key roles in their decision-making.

*Confirmatory Role*: The most common role was the confirmatory role. In this role, the RA validates the investor's initial estimate or sentiment (e.g., optimistic or pessimistic), reinforcing their confidence. Rather than prompting major revisions, the RA's advice affirms the user's original judgment. As a result, users often retain their initial estimates or make only slight adjustments. Several participants noted this alignment as a reason for not revising their predictions, stating: *"I did not change because my guess was pretty close to the AI version," "My initial guess was close enough to the algorithm's estimation that I felt more confident," "My first guess was close to the one provided by the algorithm, so I felt comfortable using it as my final guess,"* and *"I was already pessimistic and the algorithm confirming pessimism made me slightly more pessimistic."*

*Reflective Role*: In this role, the RA prompts users to reassess their initial judgments. Its forecast encourages investors to critically evaluate their reasoning, often leading them to question their optimism, pessimism, or expertise. For example: *"I have changed my estimate by making it lower because I am now thinking I might have overestimated the value of the stock,"* and *"The algorithm's estimate brought me back to reality."* One participant noted, *"The algorithm made me doubt myself, so I lowered the amount."* This reflection leads to a meaningful revision of the initial prediction, resulting in greater alignment with the RA.

*Complementary Role*: In this role, the RA enhances users' initial judgments by offering additional insights or data. Participants often described the RA as a reference or benchmark: *"I used the algorithm as a reference point to adjust my initial guess,"* and *"The algorithm provided a useful benchmark that influenced my final decision."* Others noted how the RA enriched their decision-making: *"The algorithm's prediction provided additional data that improved my initial judgment,"* and *"RA's analysis made me revise my estimate to incorporate more factual data."*

Unlike validation or reflection, this role involves integrating the RA's input to strengthen the final decision.

### 5.2.2 Integration of RA Advice

Beyond interpreting the RA's forecast, participants also varied in how they integrated its advice into their decisions. These patterns revealed four distinct user types, each reflecting a different level of trust and reliance on RA input.

*Conservative Users*: Conservative users show minimal adjustment to RA advice, preferring to stay close to their original estimates. This is evident in statements such as, *"I decided to change my value, if only by a little," "I decided to decrease my estimate a little based on the stock market prediction algorithm,"* and *"I did change my initial estimate a bit because I didn't want to be so pessimistic."* Their conservative stance suggests a preference for stability and a reluctance to place substantial trust in RA outputs, which they may view as uncertain. As a result, they make only slight changes and exhibit the lowest level of trust in RA.

*Skeptical Users*: These users cautiously integrate RA advice into their decisions. They critically evaluate RA recommendations and seek greater transparency or justification. Their cautiousness stems from concerns about the algorithm's accuracy, credibility, and ability to manage uncertainties. Participants expressed doubts such as *"RA won't be right"* and *"There is a 20% chance that it (RA) misses."* Others questioned the plausibility or boldness of forecasts, noting, *"It seems unlikely to me," "I think it is too bold,"* and *"Still, I think the decrease is a quite drastic decrease."* Nevertheless, their level of integration remains higher than that of conservative users.

*Compromised Users*: These users adopt a balanced approach, integrating both their own judgments and the RA's predictions to reach a moderate solution. They acknowledge the strengths and limitations of both human and algorithmic inputs, showing moderate trust in the RA without fully relying on it. Typically, they average their initial estimate with the RA's forecast. Participant comments such as *"I roughly estimated the average between the algorithm's prediction and my initial prediction," "A final estimate halfway between my first estimate and the algorithm might be* *the most probable,"* and *"I changed my estimation to something between the current value and the estimation of the algorithm,"* illustrate this strategy. By blending both inputs, compromised users aim to reduce the risk of over- or underestimation.

**Algorithm-aligned Users:** These users exhibit high trust in algorithmic predictions, often prioritizing them over personal judgments. They tend to adjust their decisions to closely align with the RA's outputs. Illustrative comments include: *"I trust an algorithm more than my own judgment,"* and *"I tend to trust algorithms more than people."* Statements such as *"I increased the amount to be closer to the value of the algorithm,"* and *"My best chance of getting closer to the accurate estimate is to trust the algorithm,"* reflect their active integration of algorithmic input. Their trust is further supported by a belief in the fallibility of human judgment: *"I simply trust the algorithm—humans are prone to mistakes."*

### 5.2.3 Antecedents of RA Acceptance

Thematic analysis revealed a range of antecedents related to investors and algorithms. Some serve as enablers, while others act as inhibitors. To better conceptualize these factors, we developed a 2×2 typology (Table 3), adapted from Venkatesh et al. [31]. This typology is organized along two dimensions: the source of the factor (individual vs. algorithm) and its effect on acceptance (enabler vs. inhibitor).

**Table 3**: A 2x2 Conceptual Typology of RA Antecedents

|  | Individual Factors *(User related factors)* | Algorithmic Factors *(RA related factors)* |
|---|---|---|
| Enablers *(Promote RA reliance)* | • General trust *(e.g., trusting technology without prior experience)*<br>• Domain knowledge *(e.g., lacking investment expertise)*<br>• Trust in RA owner *(e.g., confidence in a reputable financial firm)* | • Competency *(e.g., belief that RA provides accurate advice)*<br>• Robustness *(e.g., works in varied condition)*<br>• Reliability *(e.g., proven track record)* |
| Inhibitors *(Reduce RA reliance)* | • General distrust *(e.g., discomfort with automation)*<br>• Self-confidence *(e.g., prefer own judgment)*<br>• Task Familiarity *(e.g., prior experience)* | • Black-box design *(e.g., unclear logic)*<br>• Fallibility *(e.g., visible errors)*<br>• Contextual insensitivity *(e.g., ignores market sentiment or external events)* |

The first dimension, individual vs. algorithmic factors, draws on the algorithmic decision-making literature [46]. Individual factors include users' trust, confidence, experience, and reasoning, whereas algorithmic factors refer to features such as robustness, reliability, and transparency. The second dimension, enablers vs. inhibitors, is informed by dual-factor models of technology use [49]. Enablers promote reliance, while inhibitors act as barriers. Combining these two dimensions results in four distinct categories: *individual enablers, individual inhibitors, algorithmic enablers, and algorithmic inhibitors*. This typology integrates insights from both the qualitative themes and quantitative acceptance–rejection patterns measured by WOA.

***Individual enablers*** refer to personal factors that promote reliance on RA advice. A key driver is *general trust*, an inherent tendency to believe in technology or systems, often formed independently of any specific experience or observable attribute [50]. Several participants expressed this underlying tendency in the RA over their own judgment, as reflected in statements like: *"I trust an algorithm more than my own judgment," "I usually trust this kind of prediction," "I simply trust the algorithm,"* and *"I am trusting the algorithm."* Others highlighted perceived expertise, suggesting that the RA was more knowledgeable than they were. For example, one participant stated, *"The algorithm knows more about investments than me."*

*Domain knowledge* also shaped participants' reliance on RA advice. Many reported a limited understanding of financial markets, which led them to defer to the RA's estimate. This is reflected in comments such as: *"I don't have precise knowledge about the S&P 500, so I would choose to rely on a similar algorithm," "I am far from being a professional, I trust the algorithm more than myself,"* and *"I am not an expert on stocks, so I adjusted my estimate based on the algorithm's guidance."*

A third individual enabler is *trust in the RA's developer or owner*. Participants showed greater acceptance when the RA was described as being developed by a reputable financial institution. One participant noted, *"A financial investment firm is an authority in the field, capable of more accurate predictions."* Others shared similar sentiments: *"I trust the algorithm accounts for factors I'm unaware of, being designed by professionals,"* and *"An algorithm created by a financial investment firm is likely more accurate than my own predictions."* These responses suggest that institutional reputation enhances perceived credibility, promoting greater trust and adoption of RA advice.

***Individual inhibitors*** pertain to individual factors that impede RA acceptance. A common individual inhibitor is the *general distrust*, driven by negative perceptions of algorithmic decision-making. Several participants expressed skepticism about the accuracy and relevance of algorithmic forecasts. For example: *"I don't believe in any algorithms—they can't substitute for an investor's experience,"* and *"I do not trust stock market prediction algorithms."* Others questioned the fundamental limitations of predictive models: *"Like me, an algorithm cannot predict the future,"* and *"If algorithms worked perfectly, everyone would use them and make money."* Some participants also challenged the credibility of reported success rates, stating: *"These algorithms are never trustworthy; their success rates are hard to verify and often based on fine-tuned historical data."*

Another inhibitor is *self-confidence* in personal judgment. Many participants preferred to rely on their instincts or analysis rather than algorithmic input. This tendency is reflected in comments such as: *"My instinct tells me my first answer is most accurate,"* *"I believe in my prediction,"* *"I believe in my instinct,"* and *"I trust myself."* Others emphasized their reliance on analytical reasoning: *"I didn't change my estimate because I carefully checked the graph."* This sense of self-assuredness often resulted in resistance to RA advice, as participants placed greater trust in their own decision-making processes.

*Task familiarity* and prior experience also served as inhibitors of RA reliance. Some participants highlighted their long-term exposure to financial markets as a basis for trusting their own judgment over that of an algorithm. For example: *"I based my estimate on more than 25 years of watching the S&P 500. I trust myself more than any algorithm."* Others cited the use of technical analysis tools, such as Fibonacci retracement, to justify their decisions: *"Based on technical analysis, there is support around 3900 points."* This form of domain-specific expertise reinforced participants' confidence in their personal strategies, thereby limiting the influence of algorithmic input.

***Algorithmic enablers*** refer to characteristics of RAs that promote user reliance. A primary factor is *perceived competency*, which reflects users' confidence in the RA's analytical accuracy, computational power, and ability to detect meaningful patterns. Many participants expressed trust in the RA's capacity to process complex financial data and deliver more accurate forecasts than they could independently. As one participant noted: *"I trust the algorithm to analyze data more accurately than myself, especially since it is specifically built and trained to do so."* Another commented: *"I believe that the value the algorithm estimated is very reasonable and probably more accurate than mine, therefore I changed my initial estimate."* Others emphasized the RA's pattern recognition capabilities, such as: *"Algorithms follow a pattern based on past changes, and stocks have patterns that are predictable with math."*

Perceived *robustness* also emerged as a key algorithmic enabler. Participants valued the RA's capacity to incorporate a broader range of inputs than individual investors typically consider. They believed that algorithms integrate diverse variables, such as economic, social, and historical data, thereby enhancing the quality of decision-making. For example, one participant remarked, *"Algorithms analyze more than just graphs but also other social and economic factors,"* while another noted, *"I think the algorithm considers more variables than I do."* Another added, *"The algorithm must estimate based on all the 'fluctuations' of the year and predict with some accuracy."* These responses reflect a perception that RAs offer a more holistic and context-aware analytical approach.

Finally, *reliability* was frequently cited as a reason for following RA advice. Participants described the RA as a dependable tool, with its historical accuracy serving as a strong indicator of trustworthiness. Comments such as *"It is correct 80% of the time. Therefore, I decided to go with it,"* and *"If the algorithm is more accurate than professionals, I'm willing to trust it,"* illustrate how performance credibility fostered acceptance. Another participant shared, *"The algorithm's approach seemed reliable, so I slightly lowered my estimate,"* demonstrating how perceived dependability influenced adjustment behavior.

***Algorithmic inhibitors*** refer to features of the RA that discourage investor reliance. One commonly cited concern was the *black-box nature* of RA systems. Participants expressed discomfort with the lack of transparency regarding how RAs generate their predictions or what data inputs are considered. As one participant noted: *"Very little information about the algorithm. I slightly adjusted my prediction for the merit of the algorithm."* Another remarked: *"I did not change because I'm not aware of what the algorithm takes into account in order to estimate the price of the S&P 500 given today's circumstances: war, inflation."* Concerns about the algorithm's opaque computational methods were also raised, including comments like: *"More information should be given on how the algorithm calculates the estimation."*

A second inhibitor, *fallibility*, captures investor doubts about the accuracy of algorithmic forecasts. Some participants explicitly stated, *"It's not 100% accurate,"* while others expressed more general skepticism about predictive tools: *"Stock market prediction tools aren't always accurate."* This uncertainty led some to reject RA guidance, as seen in statements like: *"I did not change because I think the algorithm is a probabilistic method, but sometimes is not right,"* and *"Didn't change because the algorithm can still be wrong."*

The final inhibitor, *contextual insensitivity*, highlights perceived limitations in the RA's ability to integrate current geopolitical and economic realities. Participants questioned whether algorithms could account for external market shocks or dynamic global events. For instance, one participant remarked: *"The market is influenced by external events and emotions; the world economy is unpredictable at the moment."* Another referenced ongoing instability: *"With everything that's going on in Europe right now... I don't think the market will turn in a stable green any time soon."* Others raised concerns about reliance on outdated data, such as: *"Algorithms often use past data... but using the past months can lead to biased results since a lot happened... that will probably not happen again soon."* These perspectives illustrate how perceived rigidity and lack of situational awareness can diminish user confidence in algorithmic advice.

## 5.3 Second Quantitative Analysis and Findings

In the qualitative phase, participants described different ways of engaging with and interpreting RA roles. These patterns were categorized into three distinct RA roles (confirmatory, reflective, and complementary) and four user types (conservative, skeptical, compromised, and algorithm-aligned). To assess whether these qualitative distinctions correspond to measurable differences in reliance behavior, we conducted follow-up quantitative analysis using WOA scores of participants whose open-ended responses reflected these role or user-type classifications.

A one-way ANOVA was first conducted to examine differences in WOA across the three RA roles. The results revealed a significant effect of RA role on reliance ($F(2, 90) = 34.459$, $p<.001$, $\eta^2 = .434$), indicating a large effect size (Fig. 4). Post hoc Tukey tests revealed that participants who perceived the RA as playing a confirmatory role exhibited significantly lower reliance (WOA = 0.1810) than those who perceived it as reflective (WOA = 0.6915, $p < .001$) or complementary (WOA = 0.7466, $p < .001$). No significant difference was found between the reflective and complementary groups ($p = .864$).

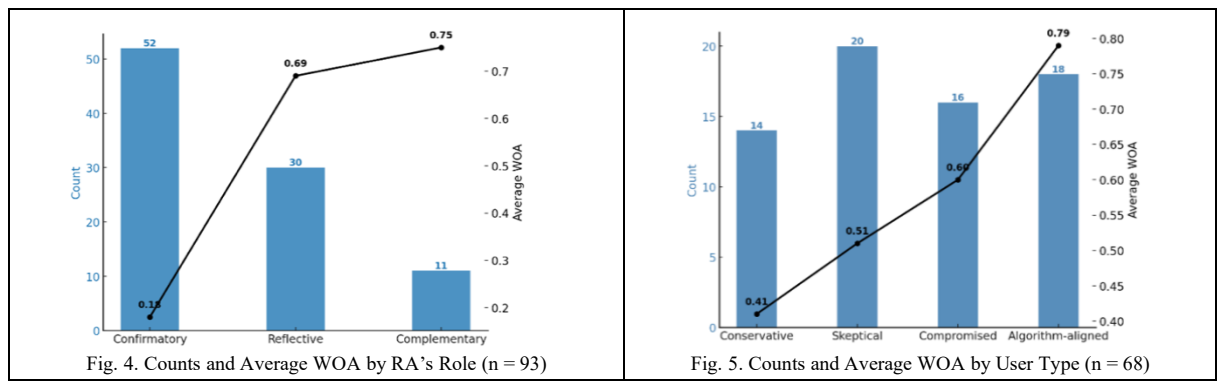

Fig. 4. Counts and Average WOA by RA's Role (n = 93)　　Fig. 5. Counts and Average WOA by User Type (n = 68)

Similarly, to examine whether reliance differed by user type, a one-way ANOVA was conducted across the four identified user types. The results showed a significant effect on WOA ($F(3, 64) = 10.899$, $p < .001$, $\eta^2 = .338$), also indicating a large effect size (Fig. 5). Post hoc Tukey tests confirmed that algorithm-aligned users exhibited the highest reliance (WOA = 0.789, $p < .001$), significantly higher than all other groups. Conservative users showed the lowest reliance (WOA = 0.409), while skeptical (WOA = 0.505) and compromised users (WOA = 0.596) fell in between. Each user type reflected statistically distinct levels of reliance, validating the classification developed in the qualitative analysis.

### 5.4 Reliability and Robustness

To ensure reliability and validity, several measures were undertaken throughout both qualitative and quantitative analyses.

#### 5.4.1 Qualitative Phase

In the thematic analysis, low-quality responses that did not align with the study objectives were excluded before coding. Additionally, responses from participants who did not follow the RA's advice were excluded from the categorization of integration styles, as they conceptually fell outside the scope of user engagement. The first two authors independently coded the data and collaboratively developed themes. Discrepancies were resolved through consensus to minimize individual bias [51]. Themes were iteratively reviewed and cross-validated against the raw data to ensure they were distinctive and exhaustive. Sufficient time was dedicated to each phase to ensure analytical depth and rigor [52]. Only recurring themes were retained in the final model.

#### 5.4.2 Quantitative Phase

*First*, in the initial experimental phase, the distribution of participants across the four experimental conditions was uneven. One group included more than 100 participants, while others had fewer than 80. Such imbalance can raise concerns about violations of the homogeneity of variance assumption, potentially undermining statistical validity. To address this, Levene's test for equality of variances was conducted. The results indicated no significant differences in group variances ($p > .05$), suggesting that the unequal sample sizes did not introduce bias into the analysis.

*Second*, in the follow-up quantitative phase, modest sample sizes (n = 93 for RA roles; n = 68 for integration styles) raised potential concerns regarding the reliability of inferential statistics. However, the analyses yielded large effect sizes, suggesting strong group differences. To further assess the robustness and stability of these effects, we conducted a bootstrapping procedure with 1,000 resamples. The bootstrapped results confirmed that the observed patterns were not due to random variation. Additionally, a post hoc power analysis using G*Power confirmed that the statistical power of both ANOVA tests exceeded the recommended threshold of 0.80 [53]. For the RA roles analysis ($\eta^2 = .434$, n = 93, 3 groups), the computed power was 1.000, and for the integration styles analysis ($\eta^2 = .338$, n = 68, 4 groups), the computed power was 0.999. These results demonstrate that the sample sizes were sufficient to detect large effects with high reliability.

*Third*, a potential confound emerged in the interpretation of the *confirmatory role*. Some participants made only minimal adjustments to their initial estimates due to a close match with the RA's forecast. This raised the question of whether such small changes reflect genuine reliance on RA advice or simply confirm pre-existing judgments without incorporating new input. To address this concern, we conducted an expert validation by six independent researchers, of whom five agreed that such minimal adjustments indicate some degree of behavioral reliance, as they involve a response to the RA's input, albeit minor. This validation supports our interpretation that the confirmatory role constitutes behavioral reliance, strengthening our thematic categorization's credibility.

*Finally*, although we adopted a Winsorization approach, we reanalyzed the data using the original (non-Winsorized) WOA values to assess whether results would differ. The findings remained consistent, with no significant differences, supporting the reliability of the analysis. Nonetheless, values outside the 0–1 range, which comprise 8% of total observations, carry behavioral significance. WOA values below 0 indicate reverse adjustment, where participants moved their estimates away from the RA's suggestion, suggesting psychological reactance, distrust, or active resistance beyond standard algorithm aversion (WOA = 0) [54]. Conversely, values above 1 reflect over-adjustment, consistent with automation bias or over-reliance on algorithmic input [55].

## 6. Discussion

This study used a multiphase mixed methods approach to examine the dynamics of human–RA collaboration. Based on our findings, we developed a conceptual framework (Fig. 6) that maps the key factors shaping user reliance on RAs. On the human side, the framework highlights four distinct user types: conservative, skeptical, compromised, and algorithm-aligned, each representing varying levels of trust and reliance. It also identifies individual-level enablers and inhibitors that shape users' willingness to rely on RA advice. On the RA side, the framework captures how users interpret the RA's role (confirmatory, reflective, complementary). It also includes RA enablers and RA inhibitors such as perceived robustness, transparency, and black-box concerns.

The framework shows that human and RA factors influence reliance both independently and through interaction. Independently, human factors such as trust or self-confidence can directly affect reliance, while RA features such as transparency or robustness can shape acceptance on their own. Interactions can occur within and across these domains. On the human side, user types such as skeptics may interact with enablers like domain knowledge, while inhibitors such as task familiarity can weaken the effect of trust in the RA's owner. On the RA side, roles, enablers, and inhibitors also interact. For example, a complementary role may be perceived as less effective when paired with contextual insensitivity, whereas reliability cues can mitigate concerns about black-box design. Across domains, users' trust or skepticism can shape how RA roles and enablers are interpreted, while RA design elements such as performance history may influence whether users adopt a conservative, skeptical, or algorithm-aligned stance. Overall, the framework shows that RA reliance is shaped by both direct effects and the interplay of human and algorithmic factors.

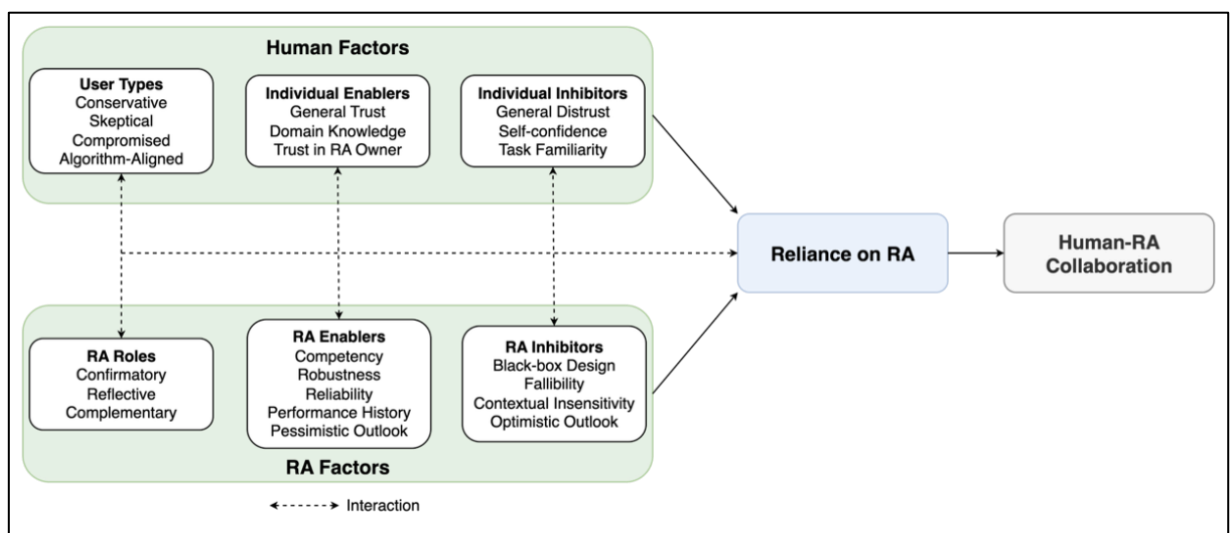

**Fig. 6.** Conceptual Framework of Factors Shaping Human-RA Collaboration

## 6.1 Implications for Research

This study makes several contributions to the Decision Support Systems (DSS) literature on human-RA collaboration, algorithmic decision-making, and financial technology adoption. First, it extends the theoretical understanding of human-RA collaboration in several ways. (i) It shifts the focus from intention-based models to actual behavioral reliance, measured through the WOA [18,33]. This shift provides a subtle understanding of how users interpret and integrate RA advice into their decision-making. (ii) This study introduces two novel roles, confirmatory and reflective, beyond the substitutive and complementary roles recognized in prior research [8,26,56]. These roles demonstrate that algorithms not only augment human actions but also validate users' existing beliefs and encourage critical thinking and reassessment of decisions. These insights provide a new perspective on understanding how RAs influence investor decisions. (iii) This study extends the innovation adoption framework established by Rogers [57], by identifying four distinct types of users specific to the context of RA usage: conservative, skeptical, compromised, and algorithm-aligned users. While Rogers primarily categorized users based on the timing of adoption—early adopters, early majority, late majority, and laggards—our classification captures varying degrees of trust in RA advice, ranging from very low to very high trust. Additionally, we broaden the work of Huang et al. [58] by moving beyond identifying criteria used to differentiate adopters from non-adopters. Instead, we identify criteria (i.e., trust levels) that distinguish among different types of adopters. This perspective offers a novel way of categorizing users, highlighting trust as a central factor in understanding user behavior. (iv) This study highlights the conceptual distinctions that are often conflated in prior research. Our findings show that trust and distrust are not merely opposite ends of a single continuum but constitute qualitatively distinct constructs, as participants expressed them in different ways [59,60]. Similarly, we distinguish between domain knowledge and task familiarity, which are often treated as overlapping. Whereas domain knowledge reflects conceptual understanding (e.g., knowing how financial markets function), task familiarity reflects experiential engagement (e.g., interpreting stock charts or making market predictions). Clarifying these distinctions helps future research better theorize the cognitive and experiential foundations of algorithmic trust and reliance. (v) This study moves beyond the traditional focus on antecedents of RA acceptance to examine user types, algorithmic roles, and acceptance factors in an integrated manner. In doing so, it also addresses a limitation of prior adoption research that typically emphasizes either enablers or inhibitors [31]. Our 2×2 typology captures both, at the individual and algorithmic levels, offering a more comprehensive perspective on the antecedents influencing RA reliance.

*Second*, this study contributes to RA design by empirically validating the impact of two key design elements—performance information and prediction outlook—on user reliance. It shows that providing performance history increases trust in RAs, while pessimistic outlooks are more likely to be trusted than optimistic ones. These findings address the mixed results in prior research. For instance, while some studies suggest that users rely more on algorithms when aware of their performance [17], others demonstrate that users discount algorithmic advice despite knowing its superior performance [21]. Similarly, while some research indicates that optimistic outlooks lead to greater reliance on algorithms [23], others argue that negative experiences are more influential than positive ones [61]. By demonstrating consistent effects across both features, our study clarifies how design features shape algorithmic trust.

*Finally,* this study uncovers insights that are transferable beyond the financial domain. For instance, transparency, trust, and framing, which shape user acceptance of RA, are equally critical in other high-stakes contexts such as healthcare, legal services, and employment decisions [62]. The trust-based user types identified here also provide a framework for designing and evaluating algorithmic systems across such settings [63]. At the same time, extending these insights to lower-stakes domains (e.g., online shopping, tourism) requires caution, as users in these contexts may be more tolerant of errors, less sensitive to transparency, and more motivated by convenience or enjoyment than trust or accuracy [64].

## 6.2 Implications for Practice

This study offers actionable insights for practitioners involved in designing, managing, and promoting RAs. *First,* the identification of three RA roles and four user types provides a framework for designing systems that adapt to user behavior. Building on this framework, an RA could infer a user's trust, confidence, and expectations from prior interactions and real-time behavioral cues (e.g., speed of following advice, requests for explanation, chart views, or the extent of adjustments) and adjust its role accordingly. For example, when a user shows low confidence in their own judgment and makes significant changes toward the RA's advice, the system could take on a reflective role, providing a brief explanation to help users make sense of the recommendation. In contrast, when a user appears highly confident and makes only small adjustments, the RA could adopt a complementary role, providing additional information to support the user's decision. Communication can also be tailored to user types. Conservative users, who are hesitant to change, may benefit from transparent explanations that build reassurance. Algorithm-aligned users, who already trust the RA, may prefer concise, data-driven recommendations. Skeptical users, who are cautious about algorithms, may be more convinced by evidence of past performance and success stories. Compromised users, who try to balance their own judgment with the RA's input, may benefit most when the system integrates their input with its own analysis to create a balanced recommendation.

*Second*, the findings on performance information and prediction framing highlight transparency as a critical factor shaping user reliance. Users often perceive RAs as "black boxes," which hinders acceptance and heightens concerns about competence, reliability, and trust. When users cannot understand how predictions are generated, they may view the RA as simplistic or unreliable. Skeptical users, in particular, become more doubtful without explanatory cues. To address these concerns, designers should incorporate transparency-enhancing features, such as explanations of prediction logic, confidence intervals, past performance metrics, and data source disclosures. These elements may reduce perceived opacity, foster trust, and support deeper user engagement.

*Finally*, the 2×2 typology of enablers and inhibitors provides a structured tool for practitioners to diagnose and address barriers to RA acceptance. For example, practitioners can leverage enablers such as prior reliability, robustness, and competency to strengthen confidence in RAs and counter general distrust. They can also mitigate inhibitors like black-box opacity by incorporating transparency-enhancing features (e.g., performance metrics and framing). Taken together, addressing both human and RA factors helps build trust, encourage adoption, and support more effective human–RA collaboration.

## 7. Limitations and Future Research

While this study provides valuable insights into human-RA collaboration, it has certain limitations that offer opportunities for future research. *First*, the study focuses on a single financial decision-making context (investment in the S&P 500). Expanding the research to other financial scenarios (e.g., retirement planning, crypto investment) or extending the insights to similar algorithmic systems, such as healthcare, education, or legal decision-making, could provide broader insights into how users interact with algorithms. *Second*, the study employs a vignette-based experimental design, which may not fully capture the complexity of real-world financial decision-making. Future studies could improve ecological validity using interactive simulations or conducting field experiments. *Third*, while this study provides a structured lens for understanding RA reliance through the conceptual framework of human and RA factors, it does not examine how these factors interact. Future research could examine such interactions. *Fourth*, future research should focus on developing validated measures for key constructs identified in this study, such as RA roles, user types, and the enablers and inhibitors, for which established measurement instruments are currently unavailable. Developing reliable and valid scales for these constructs would allow researchers to formulate hypotheses and empirically test the relationships between them. Additionally, identifying and testing strategies to overcome the identified barriers remains an important direction for enhancing RA acceptance. *Fifth*, the study lacks performance-based incentives, as participants were compensated for participation rather than prediction accuracy. This absence of real financial consequences may limit the realism of decision-making. Future research should incorporate performance-based payouts tied to accuracy to enhance the ecological validity of the findings. *Sixth*, the study found that compromised users, who claimed to average their initial estimates with the RA's advice, exhibited a slightly higher-than-expected WOA (0.60 instead of 0.50). This indicates a subtle automation bias toward algorithmic input. Future research could examine the underlying causes of this bias. *Finally,* although the second quantitative analysis examined reliance across RA roles and user types, we did not extend this to the enabler and inhibitor dimensions. Given the breadth of contextual and dispositional factors in the typology, testing all potential relationships was beyond the study's scope. We present the typology as a conceptual framework for understanding RA reliance and recommend future research to empirically validate its components and interactions.

## 8. Conclusion

This study advances the understanding of human–RA collaboration by examining how users interpret the RA's role, integrate its advice into their decisions, and what factors shape their acceptance. Guided by a multiphase mixed methods design, we investigated behavioral reliance through experimental data, uncovered interpretive and integrative patterns via thematic analysis, and validated key mechanisms through follow-up quantitative testing. The findings revealed three distinct RA roles and four user types based on varying levels of trust. Furthermore, we introduced a 2×2 typology of acceptance factors, identifying individual- and algorithm-level enablers and inhibitors. By bridging behavioral measures with qualitative reasoning, this study offers a comprehensive framework for understanding and designing more adaptive, trust-aware RA systems. It also opens new avenues for future research on algorithmic trust and collaboration.

**Appendix A: Experimental Vignette**

Imagine that you are considering investing in stock(s) contained in the S&P 500. As a part of your analysis, you now want to estimate the future value of the S&P 500 Index. The S&P represents the index value of the 500 biggest companies listed on the New York Stock Exchange. Your task is to estimate the actual value of the S&P 500 Index.

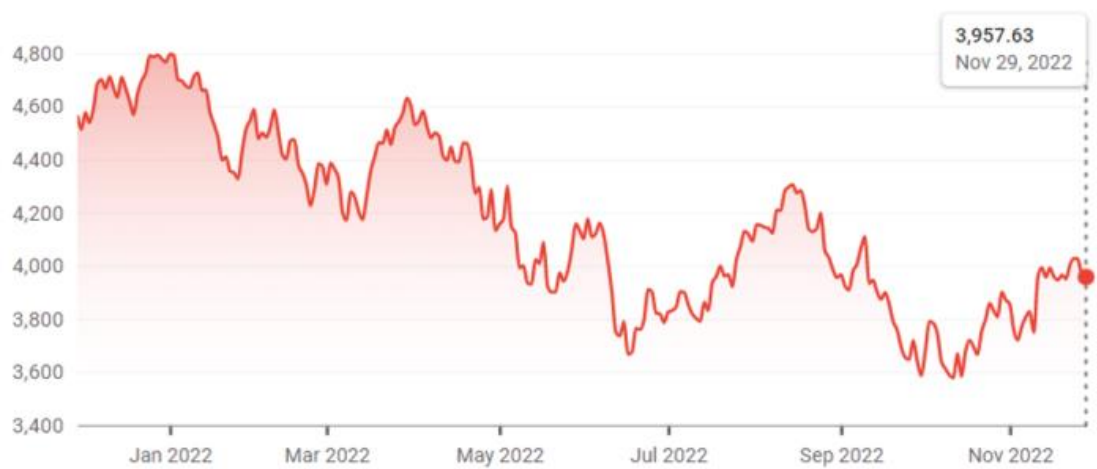

Below is a graph of the S&P 500's value over the past year.

Today's value: 3,958 points
Value at 1 year ago today: 4,567 points (Difference: 609)
Highest value over the past year: 4,797 points on January 3, 2022
Lowest value over the past year: 3,577 points on October 12, 2022
Maximum increase in points in a given month: 311 points in July 2022
Maximum decrease in points in a given month: 500 points in April 2022

Your task is to estimate the value of the S&P 500 one month from today. Please enter your estimate below.

*[After entering initial estimate, participants read one of the following]:*

[Without performance information]
A financial investment firm recently designed a stock market prediction algorithm that can estimate the future value of stocks. Considering the current situation, the algorithm is **optimistic**

**(pessimistic)** and has estimated that one month from now, the S&P 500 would be worth 4,216 (3,700) points, which is 258 points **higher (lower)** than today's value.

In light of this new information, please enter your **final** estimate below. (You may go back and see the graph details and your initial estimate if you need to.)

☐

[With performance information]
A financial investment firm recently developed a stock market prediction algorithm that can estimate the future value of stocks. By evaluating the previous performance of this algorithm, **it has been found that the estimates of the algorithm are more accurate than the average estimates of the professional investment advisors 80% of the time**.

Considering the current situation, the algorithm is **optimistic (pessimistic)** and has estimated that one month from now, the S&P 500 would be worth 4,216 (3,700) points, which is 258 points **higher (lower)** than today's value.

In light of this new information, please enter your **final** estimate below. (You may go back and see the graph details and your initial estimate if you need to.)

☐

*[After entering final estimate, all participants read]:*
Please briefly indicate why you did/did not change your initial estimate:

☐